\documentclass[12pt]{article}
\usepackage{slashed}
\usepackage{amsmath}
\usepackage{amssymb}
\usepackage{amsthm}
\usepackage{hyperref}
\usepackage{psfrag}
\usepackage{graphicx}
\usepackage{caption}
\usepackage{indentfirst}
\usepackage[utf8]{inputenc}
\usepackage{color}
\usepackage{cite}
\usepackage[usenames,dvipsnames,svgnames,table]{xcolor}
\numberwithin{equation}{section}
\hypersetup{colorlinks=true, linkcolor=blue, urlcolor=blue, citecolor=blue, linktocpage=true}

\begin{document}

\title{\vspace{-2cm} 
\begin{flushright}
{\normalsize
INR-TH-2017-014} \\
\end{flushright}
\vspace{3.5cm} 
{\bf Coset space construction and inverse Higgs phenomenon for the conformal group }}
\author{Ivan Kharuk$^{1,2,}$\footnote{E-mail address: ivan.kharuk@phystech.edu}  \\[2mm]
{\normalsize\it $^1$ Moscow Institute of Physics and
  Technology,}\\[-0.05cm]
{\normalsize\it Institutsky lane 9, Dolgoprudny, Moscow region, 141700, Russia }\\[1.5mm]
{\normalsize\it $^2$ Institute for Nuclear Research of the
Russian Academy of Sciences,}\\[-0.05cm]
{\normalsize\it 60th October Anniversary Prospect, 7a, Moscow, 117312, Russia} \\[-0.05cm]
 }
\date{}
\maketitle

\begin{abstract}

It is shown that conformally invariant theories can be obtained within the framework of the coset space construction. The corresponding technique is applicable for the construction of representations of the unbroken conformal group, as well as of a spontaneously broken one. A special role of the ``Nambu--Goldstone fields'' for special conformal transformations is clarified -- they ensure self--consistency of a theory by guaranteeing that discrete symmetries are indeed symmetries of the theory. A generalization of the developed construction to a special class of symmetry groups with a non--linear realization of its discrete elements is given. Based on these results, the usage of the inverse Higgs constraints for the conformal group undergoing spontaneous symmetry breaking is questioned. 


\end{abstract}

\vfill

\pagebreak

\hrule
\tableofcontents
\bigskip
\hrule 
\bigskip

\section{Introduction}
\label{sec:1}

The coset space technique (CST) is a very powerful tool for realizing a part of a symmetry group non--linearly\footnote{More precisely, such representations are non--homogeneous. In what follows	 those two types of realizations would not be distinguished.}. Most commonly, it is used to construct effective Lagrangians of theories undergoing spontaneous symmetry breaking (SSB) \cite{Coleman:1969sm,Callan:1969sn,Ogievetsky1974}. A less known application of this technique is the construction of Lagrangians with gauge invariance -- the latter is non--linearly realized, which allows to apply the coset space framework. For example, by properly identifying such symmetries, one can use the CST to reconstruct the Yang--Mills theory \cite{Ivanov:1976zq,Goon:2014ika} (which also provides a link to the Stueckelberg fields) and General Relativity \cite{Borisov:1974bn,Ivanov:1981wn,Goon:2014paa}. These findings naturally pose the question of whether it is possible to realize any non--linear symmetry, not only of gauge type, within the CST. 

The question of how to apply the CST to the construction of theories with non--linear discrete symmetries will be addressed in the present paper. First, it will be done on the most physically interesting example, the conformal group\footnote{The CG includes the discrete element which is realized non--linearly as the inversion operator of the coordinates of the Minkowski spacetime, $ I: ~ x^\mu \rightarrow \frac{x^\mu}{x^2} $. For convenience, both the corresponding element of the CG and its representation as the operator will be referred to as the inversion.} (CG), and then generalized to other symmetry groups. The proper interpretation of the ``Nambu--Goldsone fields'' (NGF)\footnote{They are not NGF in the conventional sense. Nevertheless, as they appear in the CST in the same manner as the usual NGF, and to have a shorthand for such fields, they will be referred to as NGF.} corresponding to non--linear spacetime symmetries is obtained by paying a careful attention to discrete symmetries of a theory. The results obtained in the paper are interesting in the following three main contexts. 

First of all, the presented technique allows to obtain conformally invariant theories within the coset space framework. Two possible approaches to the same problem were suggested in \cite{Salam:1970qk} and \cite{Wehner:2001dr}. In \cite{Salam:1970qk}, the CG was spontaneously broken and appropriate non--linear representations were obtained by starting from representations of the $ O(2,d+1) $ group. However, it is interesting to apply the CST to the unbroken CG and to obtain conformally invariant Lagrangians directly within the CST. In \cite{Wehner:2001dr}, it was suggested to double the dimensionality of the spacetime the fields of a theory live on. The resulting from such a procedure space was dubbed the biconformal space (it was also introduced in \cite{Ivanov:1981wm}). Unlike the homogeneous space of the CG, it is symmetric, which allows to apply the CST. However, this approach, to the author's knowledge, did not prove itself to be useful and is unnatural because of the doubling of the coordinates. 

One of the aims of the paper is to suggest an alternative way. The developed method allows to reproduce all consequences of the conformal invariance, such as tracelessness of the energy--momentum tensor of such theories, the transformation properties of fields, and the well--known condition that if the virial current is a divergence of some other tensor, then the theory is conformally invariant \cite{Callan:1970ze}. By paying a careful attention to the homogeneous space and the discrete element of the CG, the two--folded role of the NGF for special conformal transformations (SCT) is clarified. Namely, on the one hand, they must be introduced into a theory as fields, despite that there is no SSB. But, on the other hand, they are non--dynamical due to the fact that the solution of their equations of motion (EqM) turns out to be fixed by the symmetries. The latter requirement invokes non--trivial constraints on a theory, which ensure that the latter is not only scale invariant, but enjoys full conformal invariance. 

Secondly, the obtained results clarify a special role of the NGF corresponding to non--linearly realized spacetime symmetries (in particular, to SCT), when such symmetries become spontaneously broken\footnote{By definition, broken generators are those that have a non--trivial action on the vacuum. Such definition allows to distinguish the broken generators from a more general class of non--linearly realized ones.}. Namely, it is believed that one should use the so--called inverse Higgs constraints (IHC) \cite{Ivanov:1975zq} to express unphysical degrees of freedom (NGF for SCT) in favour of physical ones (NGF for dilations) \cite{Volkov:1973vd,Hinterbichler:2012mv}. However, the coset space construction for the CG developed in the main part of the paper questions this prescription. Namely, it implies that the NGF for SCT in the broken phase of a theory must be considered in the same way as in the unbroken one, since the solution of their EqM is still fixed by the symmetries. Thus, one is forbidden to impose any additional constraints on them, which directly contradicts the usage of IHC. To compare these approaches, an explicit example of a spontaneous breakdown of the CG is considered. It is shown that the correct effective Lagrangian can be obtained by both methods, but the usage of IHC is incompatible with the symmetries and lacks a clear physical interpretation. On the other hand, the developed approach is self--consistent and admits a clear interpretation. For the discussions on the inverse Higgs phenomenon in various contexts, the reader is referred to \cite{Nicolis:2013sga,Hidaka:2014fra,Low:2001bw,McArthur:2010zm}. A more general study of this question will be carried out by the author in the proceeding paper.

Also, by using methods unrelated to the developed construction, the spontaneous breakdown of the CG down to the Poincare subgroup is studied. It is shown that such pattern of SSB always gives rise to only one dynamical NGF corresponding to broken dilations, despite that SCT are broken as well. The underlying physics of this phenomenon is that the action of SCT on the fields of a theory is expressible in terms of dilations. This makes the introduction of the corresponding NGF unnecessary or, if introduced, redundant\footnote{This does not contradict the prescription of the developed construction to introduce the NGF for SCT because their role in the latter is qualitatively different from that of the usual NGF.} \cite{Nicolis:2013sga,Low:2001bw}. Thus, when one makes the polar decomposition, which factorizes NGF from the other fields of a theory, all dynamics of the NGF can be caught by one parameter, and, consequently, there is only one NGF. This observation indirectly supports the way the NGF for SCT are treated within the developed method and questions the usage of IHC as a proper way of excluding unphysical fields from a theory.

Finally, the coset space construction for the CG is generalized to other spacetime groups. This extends the area of applicability of the CST to groups, whose (chosen) homogeneous space is homogeneously reductive after the exclusion of all non--linear discrete symmetries. The underlying mathematical aspects of the developed method that ensure the applicability of the CST in such cases are demonstrated. As a by-product, the role of non--linear discrete symmetries in forming the algebra of the corresponding group is clarified. In particular, this allows to reveal a certain structure of the commutation relations of the algebra. 

The paper also contains an overview of the CST, which fills in the lack of literature on the connection between the method of induced representations and the CST. This connection allows to clarify the prescriptions of the CST and is important for understanding the central idea of the construction developed in the main part of the paper.

The paper is organised as follows. In section \ref{sec:2}, the review of the CST and the method of induced representations is given. Then, in section \ref{sec:3}, the technique allowing to construct conformally invariant Lagrangians and to reproduce all consequences of the conformal invariance is developed. In section \ref{sec:4}, a SSB of the CG is considered and a proper treatment of the NGF for SCT is discussed. In section \ref{sec:5}, a generalization of the presented technique to other spacetime groups is given. Finally, section \ref{sec:6} is devoted to discussions and conclusions. 

\textbf{Conventions.} Through the paper, the terms ``Lagrangian'' and ``Lagrangian density'' are used interchangeably. The ``$ \times $'' is used to show the structure of the group and does not denote the direct product of the corresponding elements. In section \ref{sec:2}, Greek ($ \mu, \nu, ... $) and Latin indices from the beginning ($ a,b, ... $) and the middle ($i,j,... $) of the alphabet are used to distinguish various elements of algebras, as well as associated elements. The signature of the metric is chosen to be $ \text{diag}(+,-,..,-) $.

\textbf{List of abbreviations:} CST -- coset space technique, SSB -- spontaneous symmetry breaking, CG -- conformal group, SCT -- special conformal transformations, NGF -- Nambu--Goldstone fields, IHC -- inverse Higgs constraints, EqM -- equation of motion, MCF -- Maurer--Cartan form, (A)dS -- (anti--)de Sitter space, CFT -- conformal field theory, DoF -- degrees of freedom. 

\section{Coset space construction}
\label{sec:2}

The application of the CST to the construction of effective Lagrangians resulting from a spontaneous breakdown of internal symmetries in the Minkowski spacetime was first developed in \cite{Coleman:1969sm,Callan:1969sn}. Soon, these results were generalized to SSB of spacetime symmetries in \cite{Ogievetsky1974}. The presented overview of this technique links to the method of induced representations \cite{Mackey:1969vt}, which clarifies the usage of the CST and explains how it can be used on other spacetime manifolds, in particular, in the (A)dS space. It is assumed that a symmetry group is global and does not possess discrete elements, the proper inclusion into consideration of which is the main goal of the paper and is done in the next section. A nice review of the original approach of \cite{Coleman:1969sm,Callan:1969sn} is done in \cite{Weinberg:1996kr}. The reader is referred to \cite{Callan:1969sn,Ivanov:1976zq,Goon:2014ika,Goon:2014paa,Weinberg:1996kr} for the application of the CST to local groups and to \cite{Goon:2014ika,deAzcarraga:1997qrt} for the construction of the Wess--Zumino terms. 

\subsection{CST and induced representations}
\label{sec:2-1}

Before considering the application of the CST to the construction of effective Lagrangians, it is useful to show that the CST is applicable in cases when there is no SSB. Namely, consider a symmetry group $ G $, its (chosen) homogeneous space $ \mathcal{A} $ with a stability group $ H $ of $ \vec{0} \in \mathcal{A} $, and a representation of $ G $ obtained by inducing a representation of $ H $. Then, the CST provides all ingredients needed to construct the most general form of the Lagrangians with the symmetry group $ G $ for the fields $ \psi(x) $ living on $ \mathcal{A} $. In this subsection, the corresponding construction is given and the notion of the induced representations is introduced.

As $ \mathcal{A} $ is the homogeneous space of $ G $, there is a one--to--one correspondence,
\begin{equation} \label{IsomHomSpace}
\mathcal{A} = G/H\;. 
\end{equation}
Denote by $ V_i $ the generators of $ H $ and let $ P_\mu $ supplement them to the full set of generators of $ G $. Then (\ref{IsomHomSpace}) establishes the isomorphism between $ \mathcal{A} $ and the orbit of $ \vec{0} $ under the action of $ g_H \in G/H $,
\begin{equation} \label{CosetInduced}
g_H = e^{iP_\mu x^\mu}\;.
\end{equation}
In particular, this makes natural to refer to $ P_\mu $ as generators of translations and to $ x^\mu $ as coordinates on $ \mathcal{A} $, which would be done from now on. Consider the left action\footnote{Equivalently, one can consider the right action of $G$ on $G/H$.} of $ G $ on $ G/H $, which for an arbitrary $ g \in G $ can be uniquely written as
\begin{equation} \label{ArbAction}
g \cdot g_H = g'_H (g,g_H) \cdot h(g,g_H)\;,
\end{equation}  
where $ h(g,g_H) \in H $ and the dot, which will be often omitted, denotes the multiplication in $ G $. This defines the transformation rule of $ g_H $ under the action of $ G $,
\begin{equation} \label{IndActCoord}
g_H \rightarrow g \cdot g_H \cdot h^{-1}(g, g_H):  ~~~  x^\mu \rightarrow x'^\mu (g, x^\mu)\;,
\end{equation}
which, as it is indicated, can be naturally interpreted as a change of coordinates. In particular, acting on $ G/H $ by $ g_1,~g_2 \in G $ successively or by $ (g_1 \cdot g_2) $ yields 
\begin{equation} \label{PropCoset}
g''_H \left( g_2, g'_H(g_1,g_H) \right) = g'_H (g_1 \cdot g_2, g_H)\,, ~~~~~ h(g_2, g'_H(g_1,g_H)) \cdot h(g_1, g_H) = h( g_2 \cdot g_1, g_H )\;,
\end{equation}
since the result of the action must be the same. 

Given the space $ \mathcal{A} $ and the action of $ G $ on its coordinates, one further wants to introduce fields living on it. This is done by using the method of induced representations. Namely, to construct the induced (from $ H $) representation of $ G $, one first introduces a space of a representation of $ H $, which will be denoted as $ \mathcal{V} $, and the corresponding representation of $ H $, $ \hat{T}(h) $, acting on it,
\begin{equation} \label{NoInd}
\hat{T}(h): ~~~ \mathcal{V} ~ \rightarrow ~ \mathcal{V}\,, ~~ \forall \, \psi \in \mathcal{V} \rightarrow \hat{T}(h)\psi \;.
\end{equation}
At this stage, $ \psi \in \mathcal{V} $ are elements of $ \mathcal{V} $, with no dependence on $ x^\mu $. As a next step, the elements of $ \mathcal{V} $ are promoted to functions with the domain $ G/H $, taking values in $ \mathcal{V} $,
\begin{equation} \label{IndSpace}
\psi \rightarrow \psi(x) \;,
\end{equation}
where it was used that each representative of $ G/H $ is uniquely determined by the values of $ x^\mu $. Finally, one defines an action of $ G $ on this space of functions to be
\begin{equation} \label{IndActonMatter}
\hat{T}(g) \psi(x) = \hat{T} \left( h^{-1} (g^{-1},g_H) \right)  \psi( x'(g^{-1}, x) )\;,
\end{equation}
where $ h $ is defined from (\ref{ArbAction}) for $ g_H $ taken at $ \vec{x} $. The obtained representation of $ G $, acting on $ x^\mu $ and $ \psi(x) $ via (\ref{IndActCoord}) and (\ref{IndActonMatter}) accordingly, is called the induced representation. In particular, (\ref{PropCoset}) guarantees that this is indeed a (non--linear) representation of $ G $. 

The simplest example illustrating the application of this technique is the construction of representations of the Poincare group from the Lorentz subgroup. First, one introduces a space of a representation of the Lorentz group, which is characterized by spin. Then, the elements of this space are promoted to dynamical fields by making them functions of $ x^\mu $, thus forming the space of the representation of the Poincare group. Finally, one defines the action of the latter on the coordinates and fields to be given by (\ref{IndActCoord}) and (\ref{IndActonMatter}) accordingly, which coincides with the usual well--known expressions. As another example, the same procedure can be applied to construct representations of the AdS group, which corresponds to inducing representation of the $ O(1,d) $ group to that of $ O(2,d) $.  

From a general perspective, induced representations are important because they naturally introduce the notion of fields and allow to obtain a representation of the full group from that of its subgroup. In particular, the initial space of a representation of $ H $ can be thought of as a set of fields defined at $ \vec{0} $ only, since at this point the induced representation of $ h \in H $, (\ref{IndActonMatter}), reduces to the initial one, (\ref{NoInd}). Then, loosely speaking, the ``induction'' of the representation corresponds to defining $ \psi $ on the whole $ \mathcal{A} $ in a way that forms a representation of $ G $.

Further, if $ G/H $ is homogeneously reductive, that is, the commutation relations of the corresponding algebra are of the form\footnote{Such decomposition of an algebra is known as the Cartan decomposition.}
\begin{equation} \label{CommutRel}
[V_i,V_j] = f_{ij}^k V_k \,, ~~~~ [P_\mu ,V_i] = f_{\mu i}^\nu P_\nu \;,
\end{equation}
where $ f_{bc}^a $ are the structure constants, the CST is applicable for the construction of $ G $--invariant Lagrangians. For that purpose, one should obtain Maurer--Cartan forms (MCF), belonging to the algebra of $ G $, defined as
\begin{equation}
g_H^{-1} d g_H = iP_\mu \omega_P^\mu + i V_i \omega_V^i \;,
\end{equation}  
where $ d $ is the differential and $ \omega_P^\mu, ~ \omega_V^i $ are the MCF for $ P_\mu $ and $ V_i $ accordingly. The transformation law of the MCF under the action of $ G $ can be found by obtaining the MCF for the transformed $ g_H $,
\begin{equation} \label{MCTransHom}
e^{-iP_\mu x'^\mu} d e^{iP_\mu x'^\mu} = (h e^{-iP_\mu x^\mu }  g^{-1}) d (g  e^{iP_\mu x^\mu}  h^{-1}) \;.
\end{equation}
By comparing the MCF in both sides of the equation above, one finds the transformation law to be given by 
\begin{equation} \label{CSTHTrans}
P_\mu \omega_P^{\mu}  ~\rightarrow~ h P_\mu \omega_P^{\mu} h^{-1}\,, ~~~~~~~~ V_i \omega_V^i ~\rightarrow~ h V_i \omega_V^i h^{-1} -i h d h^{-1}\;.
\end{equation}
Thus, the fact that $ G/H $ is homogeneously reductive guarantees that $ \omega_P^\mu $ transform homogeneously under the action of all elements of $ G $. Consequently, any $ H $--invariant combination of $ \omega^\mu_P $ would also be automatically $ G $--invariant. 

To obtain Lagrangians governing the dynamics of the fields $ \psi(x) $ constructed via the method of induced representation, which are also known as matter fields, one introduces the 1--form
\begin{equation} \label{CovarDerMatter}
D \psi = d\psi + i \hat{V}_i \omega_V^i \psi \;,
\end{equation}
where $ \hat{V}_i $ is the representation of $ V_i $ defined by the representation $ \hat{T}(h) $ of $ H $. As it follows from (\ref{CSTHTrans}), under the action of $ G $ such quantity transforms linearly, in the same way as $ \psi $ does,
\begin{equation} 
D \psi \rightarrow \hat{T} \left( h^{-1}(g^{-1},g_H) \right) D \psi \;.
\end{equation}
Consequently, it can be considered as a covariant differential of $ \psi $. The constructed forms, $ \omega_P^\mu, ~D\psi $, and $ \psi $, transform homogeneously under the action of all elements of $ G $. Thus, in the language of differential forms, $ G $--invariant Lagrangians are obtained by taking $ H $--invariant wedge products of these forms, since it automatically guarantees $ G $--invariance. Alternatively, one can define the tetrads, $ e^\mu_\sigma $, the metric, $ g_{\mu\nu} $, and the ``covariant derivatives'' of the matter fields, $ D_\mu \psi $, as
\begin{equation} \label{IngridIndRep}
\omega^\mu_P = e^\mu_\sigma dx^\sigma \,, ~~~ g_{\mu\nu} = e^{\sigma}_\mu e^{\rho}_\nu \eta_{\sigma\rho}\,, ~~~ D \psi = e^\mu_\nu D_\mu \psi dx^\nu \;,
\end{equation}
where $ \eta_{\sigma\rho} $ is the Minkowski metric, and construct $ G $--invariant Lagrangians as a $ H $--invariant combination thereof. For example, the simplest term, the invariant volume, in both approaches is given by
\begin{equation}
\mathcal{L}_{vol} = \omega^0_P \wedge \omega^1_P \wedge ... \wedge \omega^d_P = d^{d+1}x \sqrt{|g|}\;.
\end{equation}

Applied to the Poincare group, (\ref{IngridIndRep}) gives the Minkowski metric and the usual partial derivatives of matter fields. Thus, any Lorentz--invariant Lagrangian constructed from $ \psi(x) $ and $ \partial_\mu \psi(x) $ would also be automatically Poincare--invariant. More interestingly, one can apply this technique to the (A)dS space, which would give the (A)dS metric and the proper covariant derivatives of matter fields \cite{Clark:2005ht,Ortin:2015hya}.

\subsection{CST and SSB of symmetries}
\label{sec:2-2}

The logic and technique of the method of induced representations clarify the application of the CST to the construction of effective Lagrangians resulting from a spontaneous breakdown of internal or spacetime symmetries. In this case, one also starts by introducing a representation of $ H $, the stability group of a point $ \vec{0} $ of the homogeneous space $ \mathcal{A} $ of $ G $. However, now the part of $ H $ corresponding to broken symmetries is also non--linearly realized. Let $ H_0 \in H $ be a subgroup that is not spontaneously broken and, thus, is realized linearly. Then, the non--linear representation of $ H $ is obtained as the left action of $ H $ on $ H/H_0 $ \footnote{The fact that all other representations are equivalent to this one was proved in \cite{Coleman:1969sm}.}, similarly to (\ref{IndActCoord}). The corresponding coset space reads
\begin{equation}
g_{H_0} = e^{iZ_a \xi^a} \;,
\end{equation}
where $ Z_a $ are the broken generators, internal and spacetime ones. Explicitly, under the action of $ h \in H $, $ \xi^a $ transform as
\begin{equation} \label{IndNGAct}
g_{H_0} \rightarrow h \cdot g_{H_0} \cdot h_0^{-1}(h, g_{H_0}):  ~~~  \xi^a \rightarrow \xi'^a (h, g_{H_0})\;,
\end{equation}
where $ h_0 \in H_0 $ is such that $ h \cdot g_{H_0} = g'_{H_0} \cdot h_0 $. The rule above is equivalent to acting on $ g_{H_0} $ by $ h $ and then reading out the transformed $ \xi'^a $ in $ g'_{H_0} $. In particular, the action of all $ h_0 \in H_0 $ on $ \xi^a $ is linear, as it follows from 
\begin{equation}
h_0 \cdot g_{H_0} = (h_0 \cdot g_{H_0} \cdot h^{-1}_0) \cdot h_0\;.
\end{equation}
After obtaining the above representation of $ H $, one induces it to that of $ G $ by reintroducing $ \xi^a $ as functions with the domain $ \mathcal{A} $. This defines the action of $ G $ on $ G/H_0 $, corresponding to
\begin{equation} \label{SSBCoset}
g_H = e^{iP_\mu x^\mu} e^{iZ_a \xi^a(x)}
\end{equation}
coset space. Now, when $ \xi^a(x) $ are fields, they can be identified with the NGF for the broken generators. 

Further, if $ G/H_0 $ and $ H/H_0 $ are homogeneously reductive, one can apply the CST to obtain ingredients for the construction of $ G $--invariant Lagrangians. The corresponding MCF are
\begin{equation} \label{MaurerCartanForms}
g_H^{-1} d g_H = iP_\mu \omega^\mu_P + iZ_a \omega_Z^a + iV_{0i} \omega_{V_0}^i \;,
\end{equation}
where $ V_{0i} $ are the generators of $ H_0 $. By considering an analogy of (\ref{MCTransHom}) for (\ref{MaurerCartanForms}), it can be shown that $ \omega_P^\mu $ and $ \omega_Z^a $ transform homogeneously under the action of all elements of $ G $. One can also introduce matter fields, $ \psi(x) $, belonging to the (induced) space of a representation of $ H_0 $, with the associated 1--form $ D\psi $ defined via (\ref{CovarDerMatter}). Again, it can be shown that these quantities transform homogeneously under the action of $ G $. Hence, the obtained ingredients, $ \omega_P^\mu,~ \omega_Z^a, ~ \psi $, and $ D\psi $, can be used for the construction of $ G $--invariant Lagrangians, which must be a $ H_0 $--invariant wedge product thereof. Equivalently, effective Lagrangians can be obtained as $ H_0 $--invariant combinations of $ \psi $, the metric, and the covariant derivatives of NG and matter fields, $ D_\mu \xi^a $ and $ D_\mu \psi $ accordingly, defined as
\begin{equation}
g_{\mu\nu} = e^{\sigma}_\mu e^{\rho}_\nu \eta_{\sigma\rho}\,, ~~~ \omega_Z^a = e^\mu_\nu D_\mu \xi^a dx^\nu\, ,  ~~~  D \psi =  e^\mu_\nu D_\mu \psi dx^\nu \;.
\end{equation}

Summing up, the construction above demonstrates that the procedure of obtaining effective Lagrangians is fully based on the method of induced representations. It also clarifies why translations must be included into the corresponding coset space, which turns out to be of crucial importance for SSB of spacetime symmetries. However, when only internal symmetries are spontaneously broken and spacetime is good enough\footnote{By ``good enough'' it is understood that one knows how to define the proper covariant derivatives of NG and matter fields. This can always be done in the Minkowski and (A)dS spaces, which are of most interest.}, the term $ e^{iP_\mu x^\mu} $ in coset space (\ref{SSBCoset}) can be omitted, since it commutes trivially with all other elements. This reduces coset space (\ref{SSBCoset}) to the one introduced in \cite{Coleman:1969sm,Callan:1969sn}. Finally, it should be noted that for SSB of compact groups, $ H/H_0 $ is always homogeneously reductive due to the absolute antisymmetry of the structure constants.

\section{CFT and the coset space construction}
\label{sec:3}

In this section the CST will be applied to the construction of conformally invariant Lagrangians. The consideration will be restricted to the case when fields enter a Lagrangian with no more than one derivative per field\footnote{The CST allows to obtain homogeneously transforming higher derivatives terms as well \cite{Weinberg:1996kr}. The inclusion of the latter into consideration may lead to interesting consequences, which will be studied in a separate paper.}, which is enough for the purposes of the paper. Careful treatment of the homogeneous space of the CG and its discrete symmetry will provide a key for understanding a special role of the NGF for SCT. 

\subsection{The Conformal group}
\label{subsec:3-1}

An arbitrary element of the CG can be presented as a product of four basic elements\footnote{Formally, the CG also includes the reflection of coordinates. It will be shown that the presence of the latter, unlike of the inversion, does not lead to important consequences and thus will be ignored. Also, note that $ \text{Conf}(1,d) = O(2,d+1) $ as a group, not $ SO(2,d+1) $, as the latter does not include the inversion.},
\begin{equation} \label{ConfGen}
\text{Conf}(1,d) = \lbrace e^{iP_\mu a^\mu},~e^{iL_{\mu\nu}\omega^{\mu\nu}}\,,~e^{iD\sigma}\,,~I \rbrace \equiv \lbrace P \,,~ \Lambda \,,~ D \,,~ I \rbrace \;,
\end{equation}
where $ P_\mu,~ L_{\mu\nu}  $ and $ D $ are generators of translations, Lorentz transformations and dilations accordingly, $ I $ is the inversion, and $ \mu=0, .. , d  $. The CG has the following involute group automorphism, generated by the inversion:
\begin{equation} \label{CFTAuto}
G ~ \rightarrow ~ G: ~~~  \forall \, g \in G  ~ \rightarrow ~ I \, g \, I \;. 
\end{equation}
It allows to reveal the special role of the inversion in comparison with the usual discrete symmetries, such as the reflection of coordinates. Namely, under automorphism (\ref{CFTAuto}) the basic elements are mapped as
\begin{equation} \label{GroupAutoInv}
I\, e^{iP_\mu a^\mu} I = e^{iK_\mu a^\mu}\,, ~~~ I \, e^{iL_{\mu\nu}\omega^{\mu\nu}} I = e^{iL_{\mu\nu}\omega^{\mu\nu}}\,, ~~~ I \, e^{iD\sigma} I = e^{-iD\sigma}\;.
\end{equation}
The first relation in the above formula, which is a definition of SCT, states that a translation proceeded and followed by the inversion is not expressible in terms of translations, Lorentz transformations, and dilations. This fact distinguishes the inversion from the usual discrete symmetries, since the latter invoke automorphisms that map group elements to themselves (up to a sign) only. This observation will be of crucial importance in working out the proper application of the CST to the construction of conformally invariant theories in the next subsection. 

Denote by $ \mathcal{C} $ the $ (d+1) $--dimensional homogeneous space of the CG, which is known to be a pseudo--sphere, or, equivalently, the Minkowski spacetime supplemented by the causal light cone at spacetime infinity\footnote{This light cone is the boundary of the Minkowski spacetime on the corresponding Penrose diagram.}, 
\begin{equation} \label{ConfHomSpace}
\mathcal{C} = S^{1,d} = \mathcal{M}^{1,d} \oplus LC\;,
\end{equation}
where $ LC $ is the added light cone. Qualitatively, the need for adding the light cone to $ \mathcal{M}^{1,d} $ can be seen from the fact that the action of the inversion is singular at hypersurface $ x^2 = 0 $ and none of the points of $ \mathcal{M}^{1,d} $ are mapped to the origin. Thus, the action of $ \text{Conf}(1,d) $ on $ M^{1,d} $ is not transitive, and, in order to resolve this problem, the light cone at spacetime infinity must be added.

\begin{figure}[t] 
\center{\includegraphics[width=0.55\linewidth]{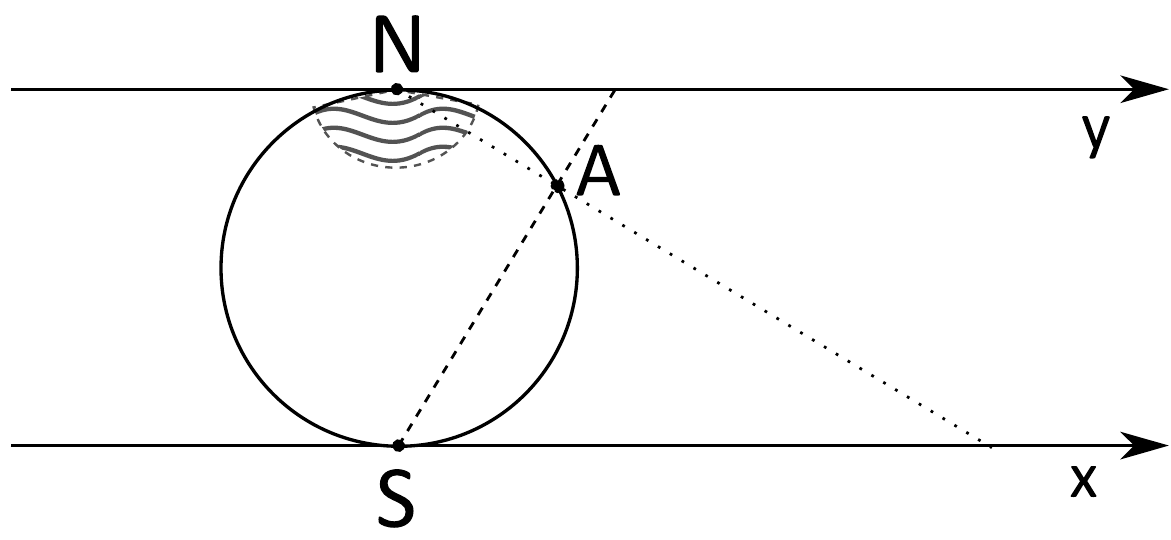}}
\captionsetup{justification=centering}
\caption{Two alternative ways of projecting a sphere to a plane, corresponding to parametrizing space by translations or SCT.} \label{TwoCoord}
\end{figure}

Identification of the stability group of the south and north poles of the pseudo--sphere, or, equivalently, of $ \vec{0} $ and of $ w^\mu = 0 \in LC $ yields, accordingly,
\begin{equation} \label{ConfHomCoset}
S: ~~ \mathcal{C} = \lbrace e^{iP_\mu x^\mu},I \rbrace\,; ~~~~~~~~
N: ~~ \mathcal{C} = \lbrace e^{iK_\nu y^\nu},I \rbrace\;,
\end{equation} 
where $ \lbrace \cdot \rbrace $ stays for a collection of elements. Thus, both translations and SCT can be chosen as ``translations'' (as defined in section \ref{sec:2-1}), since the orbit of a chosen pole under the action of the ``translations'' contains all points of $ \mathcal{C} $ except for the opposite pole. The latter is obtained by the action of the inversion, with which it is identified within the isomorphism between $ \mathcal{C} $ and coset space (\ref{ConfHomCoset}). This point is illustrated on figure \ref{TwoCoord} for the Euclidean case, when $ \mathcal{C} $ is a sphere: applying the stereographic projection from the north or the south pole of the sphere corresponds to parameterizing the space by translations or SCT accordingly, thus showing that one is free to choose any of them for the parametrization of the space. Also, this stereographic projection establishes one--to--one correspondence between the representations of $ \mathcal{C} $ as a sphere and as the Euclidean space supplemented by a point at infinity.

In what follows, the conformal algebra will be needed, which is given below:
\begin{equation} \label{CFTAlgebra}
\begin{split}
[D,P_\mu] = iP_\mu\,, ~~~~~ [D,K_\mu] = -i K_\mu\,,& ~~~~~ [K_\mu, P_\nu] = 2i (\eta_{\mu\nu}D - L_{\mu\nu})\,, \\
[K_\rho, L_{\mu\nu}] = i (\eta_{\rho\mu} K_\nu - \eta_{\rho\nu} K_\mu)\,, ~~~~&~ [P_\rho, L_{\mu\nu}] = i (\eta_{\rho\mu} P_\nu - \eta_{\rho\nu}P_\mu )\,,  \\ 
[L_{\mu\nu}, L_{\rho\sigma}] = i(\eta_{\nu\rho}L_{\mu\sigma} + \eta_{\mu\sigma}&L_{\nu\rho} -\eta_{\mu\rho}L_{\nu\sigma} - \eta_{\nu\sigma}L_{\mu\rho})\;.
\end{split}
\end{equation} 

\subsection{Coset space construction for the Conformal group}
\label{subsec:3-2}

\subsubsection{Establishing the proper coset space}
\label{subsec:3-3-1}

After clarifying the structure of the CG, it is possible to start working out the proper way of applying the CST to the construction of conformally invariant actions. As it was explained in section \ref{sec:2}, one should start with the homogeneous space of the CG. The latter is isomorphic to 
\begin{equation} \label{CosetConf1}
g_H^S = \lbrace e^{iP_\mu x^\mu},I \rbrace \;,
\end{equation}
where the upper index ``S'' emphasises that (\ref{CosetConf1}) establishes the isomorphism between $ \mathcal{C} $ and the orbit of the south pole of the pseudo-sphere under the action of (\ref{CosetConf1}). However, such coset space cannot be used in the CST framework. Indeed, the logarithmic derivative of coset element (\ref{MaurerCartanForms}), used for obtaining MCF, maps group elements to the corresponding algebra, while such transition cannot be made for discrete symmetries. At the same time, the coset space without the inversion is too small for building representations of the CG, as it does not cover the north pole of the sphere. 

Before proceeding further, a comment is of order. To avoid complications stemming from the presence of the inversion, one might try to consider representations of the conformal algebra. In this case, the corresponding coset space would be (\ref{CosetInduced}) with $ P_\mu $ given by the usual translations. Hence, the fields of such theory will be living on the standard Minkowski spacetime. Then, at first glance, it seems that the procedure of building representations of the conformal algebra can be done straightforwardly, following the construction presented in section \ref{sec:2}. However, such approach to the problem would not be self--consistent. Indeed, the problem reveals itself when one considers the action of SCT on the coordinates, which is known to be
\begin{equation} \label{SCTCoord}
x'^\mu = \frac{x^\mu + b^\mu x^2}{1 + 2b_\mu x^\mu + b^2 x^2}\;,
\end{equation}
where $ b^\mu $ are the parameters of the applied SCT and $ x'^\mu $ are the transformed coordinates. The points for which the denominator in (\ref{SCTCoord}) is zero are mapped to the infinity, and none of the points of $ \mathcal{M}^{1,d} $ are mapped back to them. Thus, one must consider $ \mathcal{C} $ as the space the fields live on, and, consequently, construct representations of the conformal group.

It is easy to understand the reason why neither (\ref{CosetInduced}) nor (\ref{CosetConf1}) coset spaces can be used for the construction of representations of the CG from a geometrical point of view. In a simple case, coset space (\ref{CosetInduced}) establishes the isomorphism between $ \mathcal{A} $ and the orbit of $ \vec{0} $, thus providing a natural coordinate chart around $ \vec{0} $. The latter forms the atlas of the space. Thus, geometrically, the proper coset space naturally endows manifold with an atlas. Adopting this logic for the case under consideration, one must consider the orbits of both poles of the sphere, since only together they span the whole $ \mathcal{C} $. One of the possible ways to do so is to use the following two cosets,
\begin{equation}
g_H^S = e^{iP_\mu x^\mu}\, , ~~~~ g_H^N = e^{iK_\nu y^\nu}\;,
\end{equation}
which provide coordinate charts around each of the poles, and apply the CST to them separately. However, this would be incorrect -- since the goal is to obtain representations of the full group, the action of all group elements on the above cosets must be well--defined. However, this does not hold for the action of the inversion, which maps each pole out of the chart. Consequently, for the CG one must use a ``two--orbit'' coset space, which acts on both poles simultaneously. As the images of the poles under the action of an element of $ G $ are two \textit{indistinguishable} points, the proper coset space is obtained by factorizing $ G $ over a subgroup leaving both poles invariant \textit{up to their exchange}. Denote by $ H_2 $ the corresponding subgroup\footnote{The fact that $ H_2 $ do form a subgroup will be proved in a more general case in section \ref{sec:5}. For the CG this fact can be checked explicitly.}, then the proper coset space is $ G/H_2 $,
\begin{equation} \label{ConfCoset}
g_H = \text{Conf}(1,d) / (SO(1,d) \times D \times I) = e^{iP_\mu x^\mu}e^{iK_\nu y^\nu}\;.
\end{equation}
Such coset space allows to cover the whole sphere by continuous group elements. Moreover, this explains why SCT, as well as translations, must be included in the coset space even when they are unbroken. In particular, a convenient way of thinking about the terms $ e^{iP_\mu x^\mu} $ and $ e^{iK_\nu y^\nu} $ in (\ref{ConfCoset}) is as of the coordinate charts around the south and north poles of the sphere accordingly\footnote{For the north pole this statement becomes precise only for $ x^\mu = 0 $, since otherwise $ e^{iP_\mu x^\mu} $ affects its image under the action of (\ref{ConfCoset}).}. Also, as it can be verified by acting by an element of the Lorentz subgroup and a dilation on (\ref{ConfCoset}), $ y^\nu $ belong to the vector representation of the Lorentz group and have the scaling dimension opposite to $ x^\mu $. Note that this observation agrees with the interpretation of $ y^\nu $ as the alternative coordinates, which correspond to the upper coordinate line on figure \ref{TwoCoord}. 

In order to avoid confusion, it should be mentioned that each chart covers the whole sphere except for the opposite pole. Thus, the sphere is not divided into two regions, one parametrized by $ x^\mu $ and the other one by $ y^\nu $. Also, this demonstrates that using $ x^\mu $ and $ y^\nu $ as coordinates simultaneously is wrong, since $ \mathcal{C} $ is a $ (d+1) $ dimensional manifold. Consequently, $ y^\nu $ in coset space (\ref{ConfCoset}) must be considered as a function of $ x^\mu $.

\subsubsection{Interpreting the NGF for SCT}
\label{subsec:3-3-2}

The geometrical picture above suggests that $ y^\nu(x) $ should provide the alternative coordinates of $ \vec{x} $. However, to answer the question of whether it is indeed the case or $ y^\nu $ is allowed to have arbitrary dynamics, like the usual NGF, strictly, one needs to approach the same construction from the perspective of induced representations. Namely, one can use the method of induced representation twice in one of the following ways,
\begin{equation} \label{RepScheme}
SG:(\psi)  \rightarrow \left[ \begin{array}{lr}
SG \times K:  \left( y^\nu, \psi(y) \right) \\ 
SG \times P:  \left( x^\mu, \psi(x) \right)
\end{array} \right.
\rightarrow  \left[
\begin{array}{lr}
\text{Conf}(1,d): \left(x^\mu, y^\nu(x), \psi(x) \right) \\
\text{Conf}(1,d): \left(y^\nu, x^\mu(y), \psi(y) \right)
\end{array} \right. \;,
\end{equation} 
where $ SG = SO(1,d) \times D $ is the scaling group, $ K = \lbrace e^{iK_\nu y^\nu } \rbrace $, arrows indicate an extension of a representation to a bigger group, and given in parentheses are the elements of the space of a representation at the corresponding stage. The final step also implies introducing the action of $ I $ as the inversion of the coordinates. Alternatively, one can induce the same representation of the $ SG $ to that of the CG directly. Then, the theorem on induction in stages \cite{Mackey:1969vt,Blattner1961}, stating that the resulting representations are equivalent, allows to establish two important consequences. The first one is that one must choose to use either $ x^\mu $ or $ y^\nu $ as coordinates, as it is done in (\ref{RepScheme}), contrary to doubling the dimensionality of the spacetime. Secondly, as both of the representations in (\ref{RepScheme}) are equivalent to the one--step one, they must be equivalent as well. This condition can be automatically fulfilled if\footnote{For convenience, the corresponding rule is given for the Euclidean case. Its generalization to the Minkowski spacetime is straightforward.}
\begin{equation} \label{Glue}
y^\nu (x) = \frac{x^\nu}{x^2}\,, ~\vec{x} \neq \vec{0}\,, ~~~~~ y(0)=N\;,
\end{equation}
and vice versa. This guarantees that $ x^\mu $ and $ y^\nu $ define the same point of $ \mathcal{C} $, and hence the theories are identical. Consequently, the dependence of $ y^\nu(x) $ on $ x^\mu $ is fixed by the symmetries, and this fact will be of crucial importance in reproducing the properties of CFT. Also, note that inducing a representation according to scheme (\ref{RepScheme}) does not lead to coset space (\ref{ConfCoset}). Indeed, in (\ref{RepScheme}) the inversion cannot be included into the intermediate step, since such set of elements would not form a subgroup. In fact, the analogous scheme for coset space (\ref{ConfCoset}) is
\begin{equation} \label{RepSchemeCoset}
SO(1,d) \times D:(\psi) ~ \rightarrow ~ SO(1,d) \times D \times I: (\psi) ~ \rightarrow ~ \text{Conf}(1,d): \left( x, y^\nu(x), \psi(x) \right)\;,
\end{equation}
where the intermediate representation is obtained by including the action of the inversion in a natural way. Again, the theorem on induction in stages guarantees that the resulting representations for schemes (\ref{RepScheme}) and (\ref{RepSchemeCoset}) must be the same. Therefore, $ y^\nu $ must be considered as a function of $ x^\mu $ and is forced to obey the gluing map of the coordinate charts, (\ref{Glue}).
 
To proceed further, one needs to obtain the MCF for coset space (\ref{ConfAltCoset}). Straightforward calculation yields
\begin{equation} \label{MKFormCFT}
\begin{split}
g_H^{-1}d g_H =& iP_\mu \omega^\mu_P + iK_\nu \omega^\nu_K + iD \omega_D + iL_{\mu\nu} \omega^{\mu\nu}_L\,, \\ 
\omega^\mu_P = dx^\mu\,, ~~~ \omega^\nu_K = dy^\nu + 2y_\rho &dx^\rho y^\nu - y^2 dx^\nu\,, ~~~ \omega_D = 2 y_\rho dx^\rho\,, ~~~ \omega^{\mu\nu}_L = -2 y^\mu dx^\nu \;.
\end{split}
\end{equation} 
Note that $ G/H_2 $ is homogeneously reductive, and thus the MCF transform homogeneously under the action of all continuous symmetries. What is more, under such transformations, due to the form of the conformal algebra, $ \omega_P^\mu $ and $ \omega_K^\nu $ do not mix with each other. The fundamental reason why the coset space $ G/H_2 $ turned out to be good (in the above sense) will be explained in section \ref{sec:5}, for now it will be left as an important observation. However, it is also required to study the transformation properties of the MCF under the action of the inversion, which is a discrete symmetry. The action of the latter on coset space (\ref{ConfCoset}) is equivalent to making group automorphism (\ref{CFTAuto}), thus leading to the following transformation of the MCF,
\begin{equation} \label{InvMKTransf}
\omega_P^\mu \rightarrow \omega_K^\mu\,, ~~~ \omega_K^\nu \rightarrow \omega_P^\nu\,, ~~~ \omega_D \rightarrow - \omega_D\,, ~~~ \omega_L^{\mu\nu} \rightarrow \omega_L^{\mu\nu}\;.
\end{equation}
Unlike the transformation rule of the MCF under the action of continuous symmetries, the action of the inversion interchanges 1--forms for translations and SCT. Conformally invariant Lagrangians, once obtained via the CST, must be invariant under such transformation, which is a new peculiarity once the discrete symmetry is included into consideration. In particular, the presence of the usual discrete symmetries, which invoke only trivial group automorphisms, would not result in the appearance of similar restrictions on the allowed combinations of the MCF. 

To understand the consequences the above requirement leads to, note that group automorphism (\ref{CFTAuto}) can also be considered as the following isomorphism of $ \mathcal{C} $ to itself,
\begin{equation} \label{HomAuto}
\mathcal{C} ~ \rightarrow ~ \mathcal{C}: ~~ \forall \, c \in \mathcal{C} ~ \rightarrow ~ \hat{I}c \;.
\end{equation}
This mapping exchanges the south and north poles of the sphere and the corresponding coordinate charts. As $ x^\mu $ and $ y^\nu $ are defined as coordinates therein, it also leads to the exchange of their roles. Then, the transformed Lagrangian can be equivalently obtained by taking the same exterior products of the MCF for the coset space
\begin{equation} \label{ConfAltCoset}
\tilde{g}_H = e^{iK_\nu y^\nu} e^{iP_\mu x^\mu}\;,
\end{equation}
but with $ \omega_P^\mu $ and $ \omega_K^\nu $ exchanged, as now $ y^\nu $ are used as the coordinates of $ \mathcal{C} $. This transformation must be a symmetry, which can be fulfilled only if the new translational MCF, $ \omega_K^\nu = d y^\nu $, are the pullbacks of the old ones, $ \omega_P^\mu = dx^\mu $, after change of coordinates (\ref{HomAuto}). This forces $ y^\nu(x) $ to be given by (\ref{Glue}), thus reproducing the result obtained by the method of induced representations. Another argument for fixing $ y^\nu $ is that, as there is no SSB, there is no reason to assume the existence of dynamical fields in a theory \cite{Salam:1969rq}. This also agrees with the fact that CFT do not necessarily possess such a field. 

To put in other words, the argumentation used above and in the method of induced representations is based on the following fact. One is free to choose either $ x^\mu $ or $ y^\nu $ as coordinates on $ \mathcal{C} $. In the CST framework, this leads to the usage of the coset spaces (\ref{ConfCoset}) or (\ref{ConfAltCoset}) accordingly. The constructed Lagrangians cannot depend on the choice of coordinates, which is possible only if $ x^\mu $ and $ y^\nu $ are connected via gluing map (\ref{Glue}).  
 
\subsubsection{Reproducing transformation properties}
\label{subsec:3-3-3} 
 
Before obtaining conformally invariant Lagrangians within the CST, it should be verified that the latter correctly reproduces the transformation properties of the coordinates and fields under the action of the CG. According to scheme (\ref{RepSchemeCoset}), matter fields should be introduced as representations of the $ SO(1,d) \times D $ group, and, consequently, are characterized by their spin and scaling dimension. In particular, this is in agreement with the fact that fields in CFT do not carry an index characterising their charge under SCT. As the Lorentz group and dilations are realized linearly, it is only needed to verify the action of SCT on the coordinates and fields. To do that, one should bring the product of $ e^{iK_\mu b^\mu} $ and $ g_H $ to the standard form,
\begin{equation} \label{SCTAction}
e^{iK_\nu b^\nu} g_H = e^{iP_\mu x'^\mu} e^{iK_\nu y'^\nu} e^{iD\sigma(b,x)} e^{iL_{\mu\nu}\omega^{\mu\nu}(b,x)}\;. 
\end{equation} 
According to (\ref{IndActonMatter}), this implies that matter fields transform as
\begin{equation} \label{MattTransfSCT}
\psi(x) \rightarrow Rep( e^{-iD\sigma(x,-b)},~ e^{-iL_{\mu\nu}\omega^{\mu\nu}(x,-b)} )\psi(x)\;,
\end{equation}
where $ Rep(\cdot) $ is a representation of $ SO(1,d) \times D $ appropriate for $ \psi $. Note that, due to the commutation relations of the conformal algebra, $ \sigma(b,x) $ and $ \omega^{\mu\nu}(b,x) $ do not depend on $ y^\nu $. The transformation law of the coordinates can be found as follows,
\begin{equation}
I \cdot e^{iP_\nu b^\nu} \cdot (Ie^{iP_\mu x^\mu}) = e^{iP_\mu x'^\mu} \cdot e^{iK_\nu y'^\nu} e^{iD\sigma(b,x)} e^{iL_{\mu\nu}\omega^{\mu\nu}(b,x)}  ~~ \Rightarrow ~~ \frac{x'^\mu}{x'^2} = \frac{x^\mu}{x^2} + b^\mu \;,
\end{equation} 
where the explicit form of SCT was used. The infinitesimal version of this transformation can also be obtained directly by commuting the SCT and translations in the left-hand side of (\ref{SCTAction}). The infinitesimal versions of $ \sigma $ and $ \omega_{\mu\nu} $ can be easily found to be
\begin{equation}
\sigma = 2b_\mu x^\mu \,, ~~~ \omega^{\mu\nu} = b^\mu x^\nu - b^\nu x^\mu \;,
\end{equation}
which coincide with the standard well--known expressions. Then, the group property guarantees that they will coincide at the non--linear level as well. Thus, the CST correctly reproduces the transformation properties of fields under the action of SCT.

\subsubsection{Constructing conformally invariant Lagrangians}
\label{subsec:3-3-4}

Now everything is prepared for the construction of conformally invariant Lagrangians in the coset space framework. The 1--form associated with matter fields via rule (\ref{CovarDerMatter}) is
\begin{equation} \label{CovarDerMatterFields}
D\psi = \partial_\mu\psi dx^\mu + 2y^\nu ( \eta_{\mu\nu}\Delta + i \hat{S}_{\mu\nu})\psi dx^\mu \equiv ( \partial_\mu\psi + y^\nu \hat{N}_{\nu\mu} \psi ) dx^\mu\;,
\end{equation}
where $ \Delta $ and $ \hat{S}_{\mu\nu} $ are appropriate representations of the dilation and Lorentz groups. Conformally invariant Lagrangians are then obtained by combining $ D_\mu \psi,~ \psi,~ \omega_P^\mu$, and $ \omega_K^\nu $ in a $ (SO(1,d) \times D) $--invariant way. 

The construction of conformally invariant theories will go from the simplest case to the most general one in three steps. First, consider the case when there are no matter fields. Then, the Lagrangian is a pure ``kinetic term'' for $ y^\nu $,
\begin{equation}
\mathcal{L}_y = \mathcal{L}_{kin}(\omega_P^\mu, \omega_K^\nu)\;,
\end{equation}
where $ \mathcal{L}_{kin}(\omega_P^\mu, \omega_K^\nu) $ is an arbitrary function constructed as a $ (SO(1,d) \times D) $--invariant wedge product of $ \omega_K^\nu $ and $ \omega_P^\mu $. As the variation of the Lagrangian with respect to $ y^\nu $ is proportional to $ \omega^\nu_K $, the EqM of $ y^\nu $ always admit the following solutions\footnote{Of course, there can be other solutions. For example, in $ (d+1)=4 $ dimensions one of the possible kinetic terms for $ y^\nu $ is $ ( \partial_\mu y_\nu - \partial_\nu y_\mu )^2 $, which is free of instabilities and admits a lot of solutions.},
\begin{equation} \label{MCFConstr}
\omega_K^\nu = 0 ~~~ \Rightarrow ~~~ 
y^\nu = 0 ~~  \cup ~~ y^\nu = \frac{x^\nu}{x^2}\;.
\end{equation}
The second one provides the gluing map between the coordinate charts, (\ref{Glue}). As both of them are obtained by setting the homogeneously transforming form $ \omega_K^\nu $ to zero and, hence, are compatible with all symmetries, none of them is preferable on the level of EqM. However, as it was explained above, one should choose the second solution among all possible ones.

In particular, the reason why conditions $ \omega_K^\nu = 0 $ give the gluing map is that solution (\ref{Glue}) is the only one that is compatible with all symmetries, including the inversion. Thus, it must correspond to setting some homogeneously transforming quantity to zero, and the 1--form $ \omega_K^\nu $ is the only candidate for such a role.

As a second step, consider the case when matter fields mix with $ y^\nu $ via their covariant derivatives only. Denote by $ \mathcal{L}_\psi $ the Lagrangian for the matter fields. Then varying the action with respect to $ y^\rho $ yields,
\begin{equation} \label{VarySCT}
\frac{\delta L(\psi, D_\mu \psi)}{\delta y^\rho} + \frac{\delta \mathcal{L}_{kin}(\omega_K^\nu)}{\delta y^\rho} \equiv V_\rho + \frac{\delta\mathcal{L}_{kin}}{\delta y^\rho} = 0\;,
\end{equation}
where $ V_\rho $ is the virial current,
\begin{equation} \label{VirialCurrent}
V_\rho = \frac{\delta L}{\delta \partial_\mu \psi} ( \Delta \eta_{\mu\rho} + i \hat{S}_{\mu\rho} ) \psi \equiv V^{(0)}_\rho + V^{(1)}_{\rho\nu} y^\nu\;,
\end{equation}
where $ V^{(0)}_\rho $ and $ V^{(1)} _{\rho\nu}$ do not depend on $ y^\nu $. Requiring gluing map (\ref{Glue}) to be a solution of these EqM forces the virial current to vanish identically. Thus, the class of CFT with zero virial current (and, consequently, with a traceless energy--momentum tensor \cite{Nakayama:2013is}) corresponds to such class of Lagrangians. 

For example, this result reproduces the well--known facts that free massless spin--0, spin--1 and spin--$\frac{1}{2}$ field theories with the canonical kinetic terms are conformally invariant in $(1+1)$, $(3+1)$, and in arbitrary number of dimensions accordingly (only in these cases the corresponding virial currents are zero). To provide an example of how the construction above works, consider an arbitrary integer--spin field $ \psi_a $. Assuming a quadratic kinetic term, the relevant part of the Lagrangian is
\begin{equation} \label{ExplCalcConstr}
\mathcal{L} = \frac{1}{2} C^{\mu a \nu b }(D_\mu \psi_a)(D_\nu \psi_b) = \frac{1}{2} C^{\mu a \nu b} \big( \partial_\mu \psi_a \partial_\nu \psi_b + 2\partial_\mu \psi_a y^\sigma \hat{N}_{\sigma\nu} \psi_b + y^\sigma \hat{N}_{\sigma\mu} \psi_a y^\rho \hat{N}_{\rho\nu} \psi_b \big)\;,
\end{equation}
where $ C^{\mu a \nu b} $ is some constant tensor, symmetric in $ (\mu a) \leftrightarrow (\nu b) $. Condition (\ref{VarySCT}) then yields
\begin{equation} \label{ExplVarySCT}
C^{\mu a \nu b}  (\partial_\mu \psi_a + y^\sigma \hat{N}_{\sigma\mu} \psi_a ) \hat{N}_{\rho\nu} \psi_b = 0 ~~~ \Rightarrow ~~~ C^{\mu a \nu b} \hat{N}_{\rho\nu} \psi_b = 0\;.
\end{equation}
As it can be explicitly checked, (\ref{ExplVarySCT}) holds for the mentioned above fields and dimensionalities of the spacetime. Importantly, after taking into account this constraint, only the first term in Lagrangian (\ref{ExplCalcConstr}) does not vanish. That is, if (\ref{VarySCT}) is fulfilled, the covariant derivatives of matter fields simplify to the usual ones on the Lagrangian level. In particular, this observation explains why scale and conformally invariant Lagrangians look the same, yet conformal invariance requires usage of covariant derivatives (\ref{CovarDerMatterFields}) and implies (\ref{VarySCT}).

Finally, the most general class of conformally--invariant Lagrangians is obtained by allowing matter fields to mix with $ \omega_K^\nu $ directly. At first glance, it may seem impossible to satisfy conditions (\ref{Glue}), since the presence of the interaction terms does not allow to fix a solution of $ y^\nu $'s EqM. However, it turns out that it is possible to organise the interaction terms between $ \psi $ and $ y^\nu $ to sum up to a total derivative, which automatically guarantees that (\ref{Glue}) is a solution. Clearly enough, such situation can take place only if $ V^{(0)}_{\rho} $ is a total derivative, as only in this case the linear in $ y^\nu $ terms can sum up to a total derivative after adding an appropriate term with $ \omega_K^\nu $. As a straightforward but lengthy calculation, presented in the appendix, demonstrates, this is also a sufficient condition. That is, if the virial current is a total derivative, the following Lagrangian contains the interaction terms only via full derivative,
\begin{equation} \label{OnlyViaDer}
V^{(0)}_\rho = \partial_\mu L^\mu_\rho, ~~~~ \mathcal{L} = \frac{1}{2}  D\psi \wedge \star D\psi + \epsilon_{\mu_0 ... \mu_d} L^{\mu_0}_\nu \omega_K^\nu \wedge \omega_P^{\mu_1} \wedge ... \wedge \omega_P^{\mu_d} \;,
\end{equation} 
where $ \star $ is the Hodge dual operator and the kinetic term for $ y^\nu $ was omitted. Thus, a class of scale--invariant theories, which are, in fact, conformally--invariant after an improvement of the energy--momentum tensor \cite{Callan:1970ze}, correspond to the Lagrangians of type (\ref{OnlyViaDer}). This result is also known in another context \cite{Volkov:1973vd}. In particular, this class of Lagrangians describes massless spin--0 fields in arbitrary dimensionality of the spacetime and the so--called ``elastic'' vector field theory \cite{Riva:2005gd,ElShowk:2011gz}, the virial currents of which are total derivatives.

Note that Lagrangian (\ref{OnlyViaDer}) can be split into two parts,
\begin{equation} \label{WessZumino}
\mathcal{L} = \frac{1}{2} d\psi \wedge \star d \psi + d\tilde{\mathcal{L}}(y,\psi)\;.
\end{equation}
The second term in the expression above is a total derivative and, consequently, can be dropped without affecting the dynamics of a theory. Then, the first term alone can be considered as a special type of the Wess--Zumino term that can arise on the manifolds whose atlas must contain more than one coordinate chart. Examples of such terms are given by the standard Lagrangians for the massless spin--0 and elastic vector field theories, which are obtained by dropping the corresponding total derivative part from the complete Lagrangian (\ref{OnlyViaDer}). 

Thus, the developed technique reproduces all consequences of the conformal invariance and allows to obtain the Lagrangians of known CFT. The crucial point that has been overlooked so far is that for the inversion to be a symmetry of a theory, the NGF for SCT must be introduced into the corresponding theory. Moreover, a solution of EqM of these NGF is fixed by the symmetries to be given by gluing map (\ref{Glue}), which leads to collective constraints (\ref{VarySCT}) on the theory. The role of the NGF for SCT in a spontaneously broken phase will be discussed in the next section.

As a final comment, it should be mentioned that imposing the constraints
\begin{equation}
\omega_K^\nu = 0
\end{equation}
as a way of fixing the dependence of $ y^\nu $ on $ x^\mu $ from the very beginning is wrong. Indeed, setting $ \omega^\nu_K $ to zero would not allow to reproduce CFT described by the Lagrangians of type (\ref{OnlyViaDer}), thus a posteriori demonstrating a failure of this approach.

\section{Connection with the inverse Higgs phenomenon}
\label{sec:4}

\subsection{Comparison with the inverse Higgs phenomenon}
\label{subsec:4-1}

According to the presented construction, the NGF for SCT play the special role in conformally invariant theories. Interestingly, this allows to question the inverse Higgs phenomenon for CFT undergoing SSB. To illustrate this, suppose that the SSB pattern is
\begin{equation} \label{ConfToPoincare}
\text{Conf}(1,d) \rightarrow ISO(1,d)\;,
\end{equation}
so that the dilations and SCT become broken. Following the logic of the standard construction, one introduces the NGF $ \pi(x) $ and $ y^\nu(x) $ for the broken dilations and SCT accordingly. Then, since $ [P_\mu, K_\nu] \sim \eta_{\mu\nu}D $, one can impose IHC \cite{Hinterbichler:2012mv}
\begin{equation} \label{IHCConf}
\omega_D = 0 ~~~ \Rightarrow ~~~ y_\nu (x) = -\frac{1}{2}e^{-\pi(x)}\partial_\nu \pi(x)\;,
\end{equation}
where $ \omega_D $ is the MCF for the dilation for pattern (\ref{ConfToPoincare}). The expression above is compatible with all symmetries, and thus allows to express the unphysical degrees of freedom, $ y^\nu $, in term of the physical one, $ \pi $, the dilaton, in an invariant way. Following the prescriptions of IHC, one further substitutes (\ref{IHCConf}) into $ \omega_K^\nu $ and $ \omega_P^\mu $ and constructs effective Lagrangians as a $ SO(1,d) $--invariant wedge product thereof. 

However, the construction above does not respect all of the symmetries. To see this, note that the symmetry considerations from the previous section are valid for a spontaneously broken CG as well. Namely, as the inversion is a symmetry of a theory, $ y^\nu $ is still forced to obey the gluing map between the coordinate charts around the north and south poles of the sphere. On the other hand, IHC (\ref{IHCConf}) suggest excluding $ y^\nu $ in favour of the dilation field. This explicitly breaks the condition that $ y^\nu $ must obey gluing map (\ref{Glue}) and, hence, contradicts the symmetry requirements. 

The other way to question the usage of IHC is to note that, as it was explained in the previous section, one is free to choose either $ x^\mu $ or $ y^\nu $ for the parametrization of the spacetime. Consequently, they can be used interchangeably within the CST. However, IHC (\ref{IHCConf}) breaks this condition. To make this statement more manifest, suppose that $ y^\nu $ were chosen as the coordinates of the spacetime. Then, from the IHC for this case follows
\begin{equation} \label{IHCAlt}
x_\mu = \frac{1}{2} e^{-\pi}\partial_\mu \pi\;.
\end{equation}
As $ x^\mu $ can be still thought of as coordinates, it is hard to assign a meaningful interpretation to conditions (\ref{IHCAlt}), which makes the usage of IHC doubtful. 

The author believes that the correct usage of the CST for a spontaneously broken CG is the following. In an unbroken phase of a theory, $ y^\nu (x) $ provide the gluing map between the coordinate charts, and the EqM of $ y^\nu $ invoke constraints (\ref{VarySCT}) ensuring that the virial current is zero or a total derivative. This must hold for any CFT, including a spontaneously broken one, since SSB cannot lead to the violation of this condition. Indeed, if the virial current was a total derivative in terms of the UV fields, it remains so in the IR, as, formally, SSB corresponds to a change of field variables. Thus, the EqM of $ y^\nu $ must again be considered as additional constraints on a theory, which distinguish the class of effective Lagrangians with a non--linear realization of the conformal symmetry. 

In particular, since the fields in CFT do not carry an index characterizing their charge under SCT, the suggested approach agrees with works \cite{Nicolis:2013sga,Low:2001bw,Watanabe:2013iia}. Therein it was pointed out that if some of the broken generators do not produce independent fluctuations of an order parameter, then the corresponding NGF are redundant and should not be introduced at all \cite{Watanabe:2013iia}. In particular, by using another approach, in \cite{Hidaka:2014fra} it was argued that the breakdown of SCT never invokes the corresponding NGF.

To demonstrate that the technique explained above works, consider two spin--0 fields in the Minkowski spacetime of dimension  4 governed by the Lagrangian
\begin{equation} \label{LagrSSB}
\mathcal{L} = \frac{1}{2} ( \partial_\mu \varphi )^2 + \frac{1}{2} ( \partial_\mu \xi )^2 - \lambda ( \varphi^2 - \xi^2 )^2\;.
\end{equation}
Such theory admits the following SSB pattern,
\begin{equation} \label{SSBPattern}
\varphi(x) = \xi(x) = C\, , ~~~~~ \text{Conf}(1,d) \rightarrow ISO(1,d)\;,
\end{equation}
where $ C $ is some non--zero constant. The initial Lagrangian contains two DoF, one of which becomes a radial mode after the SSB, while the second one corresponds to a NGF. Thus, there is only one NGF, and one has to find a way of eliminating the unphysical NGF for SCT from the theory in the CST framework. Before performing calculations within the CST, it is useful to obtain the corresponding effective Lagrangian explicitly. The fluctuations of the fields satisfying the $ \sigma $--model constraint $ \varphi^2 = \xi^2 $ can be parametrized as
\begin{equation} \label{Fluct}
\varphi (x) = e^{\pi(x)} C\,, ~~~ \xi (x) = e^{\pi(x)} C\;,
\end{equation}
where $ \pi $ is the NGF. Note that this parametrization coincides with the parametrization of the NGF for the broken dilations in the coset space framework. Substituting this back into Lagrangian yields
\begin{equation} \label{ExplSigma}
\mathcal{L}_{eff} = C^2 e^{2\pi} ( \partial_\mu \pi)^2\;.
\end{equation}
The same effective Lagrangian can also be obtained within the machinery of the developed technique. The corresponding coset space reads
\begin{equation} \label{CosetCorrect}
g_H = e^{iP_\mu x^\mu} e^{iK_\nu y^\nu} e^{i D\pi}\;.
\end{equation}
Note the order of the SCT and dilaton in this coset space. It ensures that gluing map (\ref{Glue}) is always a solution for $ y^\nu $ alone and reflects the fact that one should build non--linear representations of $ SO(1,d) \times D $ first. The corresponding MCF are
\begin{equation} 
\omega^\mu_P = e^{\pi} dx^\mu\,, ~~~ \omega^\nu_K = e^{-\pi} ( dy^\nu + 2y_\rho dx^\rho y^\nu - y^2 dx^\nu )\,, ~~~ \omega_D = 2y_\rho dx^\rho + d\pi\;,
\end{equation}
and the form of $ \omega_L^{\mu\nu} $ is irrelevant for present purposes. In $ (d+1) = 4 $ all conformally invariant Lagrangians for spin--0 fields are of type (\ref{OnlyViaDer}). Note that $ \omega_D $ is formally equivalent to the term $ D\psi $ for scalars in an unbroken phase. Therefore, replacing $ D\psi $ by $ \omega_D $ in Lagrangian (\ref{OnlyViaDer}) does not violate the condition that the interaction terms between $ y^\nu $ and $ \pi $ sum up to a total derivative. Then, (\ref{ExplSigma}) is the first term in (\ref{WessZumino}), with the corresponding $ L_\nu^\mu = \delta^\mu_\nu $ \footnote{Unlike the construction in the Appendix, the factor $ e^{2\pi(x)} $ in the second part of Lagrangian (\ref{OnlyViaDer}) is automatically reproduced by the wedge product of the MCF $ \omega_K^\nu $ and $ \omega_P^\mu $.}. Thus, the suggested approach allows to reproduce Lagrangian (\ref{ExplSigma}). Also, it is easy to see that the effective Lagrangian can be obtained by writing the initial one in the form similar to (\ref{OnlyViaDer}) and then substituting fluctuations (\ref{Fluct}) therein.

Effective Lagrangian (\ref{ExplSigma}) can also be reproduced by the means of the IHC method. For the case under consideration, the corresponding constraint reads
\begin{equation} \label{IHCScalar}
\omega_D = 0 ~~ \Rightarrow ~~ y_\nu(x) = - \frac{1}{2} \partial_\nu \pi(x)\;.
\end{equation}
Then, it is easy to check that effective Lagrangian (\ref{ExplSigma}), up to a constant multiplier and a total derivative term, is
\begin{equation} \label{LagrIHC}
\mathcal{L}_{IHC} = D_\mu y^\mu\;,
\end{equation} 
where $ D_\mu y^\nu $ is the covariant derivative of $ y^\nu(x) $. 

Considered formally, Lagrangian (\ref{IHCScalar}) describes a massless scalar field and hence is conformally invariant. However, a closer look allows to establish that it is not so. Indeed, the above Lagrangian can be rewritten as the wedge product of the MCF, which have transformation properties (\ref{GroupAutoInv}) under the action of the inversion. According to (\ref{IHCScalar}), $ y^\nu $ no longer provide the gluing map of the coordinate charts, and hence the inversion is not a symmetry of the theory. Consequently, Lagrangian (\ref{LagrIHC}) is not conformally invariant and imposing IHC is not the correct way of obtaining effective Lagrangians resulting from SSB of the CG.

Thus, the technique developed in the previous section is applicable for a spontaneously broken CG as well. Moreover, unlike the inverse Higgs phenomenon, the way it treats the NGF for SCT admits clear physical and mathematical interpretations.

\subsection{Polar decomposition}
\label{subsec:4-2}

The usage of the polar decomposition allows to provide one more argument in support of the point of view that the breakdown of SCT does not lead to the existence of the corresponding NGF. By definition, a polar decomposition is the factorization of NGF and matter fields \cite{Weinberg:1996kr}, 
\begin{equation} \label{PolarDecomp}
\psi(x) = \gamma(x) \tilde{\psi}(x)\;,
\end{equation}
where $ \psi(x) $ stands for all fields of a theory and $ \tilde{\psi}(x) $ is such that it does not include NGF. Usually, the polar decomposition is used in the case of SSB of internal symmetries, corresponding to some pattern $ G \rightarrow H $. Based on compactness of $ G $, it is possible to show that $ \gamma(x) $ always exists and, at a given $ \vec{x} $, is a representative of $ G/H $. In fact, the same trick can be done for the SSB pattern
\begin{equation} \label{PatternCl}
\text{Conf}(1,d) \rightarrow ISO(1,d)
\end{equation}
as well. Indeed, since the Poincare group is not broken, an order parameter can be formed only by a set of spin--0 fields. Then, without loss of generality, further consideration will be restricted to the theory described by Lagrangian (\ref{LagrSSB}), which is simple yet catches all main features of such theories. The polar decomposition then takes the form
\begin{equation} 
\left( \begin{array}{lr}
\xi(x) \\ \varphi(x) 
\end{array} \right)
= C(x) u_\alpha(x) \;,
\end{equation} 
where $ u_\alpha $ is a unit norm vector. Such decomposition preserves the number of DoF, and, as it can be seen by comparing (\ref{Fluct}) and (\ref{PolarDecomp}), $ C(x) $ corresponds to the broken dilations. This illustrates that the dynamics of all NGF can be caught by one parameter from the very beginning, and, consequently, there is no need in introducing the NGF for the broken SCT. This can also be explained in terms of an action of the generators on the order parameter as follows. The action of SCT on the fields is expressible in terms of a dilation, hence the corresponding fluctuation is already included into $ C(x) $. In other words, SCT do not give rise to dynamical NGF because they do not have their ``own'' action on the fields. 

Another way to arrive at the same conclusion is the following. Since the Poincare group is unbroken, the derivation of the usual Nambu--Goldstone theorem goes with no changes up to concluding that the number of NGF equals the number of the broken generators. A general rule for counting the number of NGF, following from the Nambu--Goldstone theorem, is that there are as many massless NGF as there are independent actions of the broken generators on an order parameter. In the case of SSB of internal symmetries, all of them are independent, and thus the number of NGF equals the number of the broken generators. For SSB of spacetime symmetries, however, not all of them are necessarily independent \cite{Low:2001bw,Watanabe:2013iia}, but the rule is still applicable. For pattern (\ref{PatternCl}), the only independent action, as it was explained above, is that of the dilation. Consequently, there is only one dynamical NGF, and other NGF should not be introduced at all rather than excluded via IHC. 

\section{Generalization}
\label{sec:5}

In this section the construction presented in section \ref{sec:3} for the CG will be generalized to other suitable spacetime groups. It will be shown that the correct usage of the CST ensures that the appropriate coset space is homogeneously reductive, and thus the CST is applicable for the construction of invariant Lagrangians. This part of the paper is mostly mathematical and is aimed at displaying the underlying mathematical aspects of the developed method and to prove some aspects that were left as observations in section \ref{sec:3}.

The general set--up of the problem is the following. Let $G$ be a symmetry group and $ A^d_{iso} $ be its $d$--dimensional homogeneous space with a stability group $ \tilde{H} $ of a point $ \vec{0} \in A^d_{iso} $. Then, there is the isomorphism
\begin{equation} \label{IsomGen}
A^d_{iso}=G/ \tilde{H}\;.
\end{equation} 
Note that for a given $ G $ there are various homogeneous spaces. For example, the $ O(2,d+1) $ group can be realized as the $ AdS $ space, $ AdS_{1,d+1}= O(2,d+1) / O(1,d+1) $, and as the conformal group in $ (d+1) $--dimensions, corresponding to the $ O(2,d+1) / (O(1,d+1) \times D) $ coset space. Isomorphism (\ref{IsomGen}) suggest identifying an element $ g_{\tilde{H}} \in G/\tilde{H} $ with the point of $ A^d_{iso} $ obtained by the action of the former on the origin. In general case, $ G/\tilde{H} $ consists of continuous and discrete elements, which will be denoted as $ e^{iP_\mu x^\mu} $ and $ T_m, ~m=1,...n\,, $ accordingly. Then, the discrete elements are identified with a finite set of points $ \lbrace z_m \rbrace $, while $ e^{iP_\mu x^\mu} $ is isomorphic to $ A^d_{iso} \setminus \lbrace z_m \rbrace $. The latter fact makes natural to refer to $ P_\mu $ as generators of translations and to $ x^\mu $ as coordinates on $ A^d_{iso} $. 

Denote by $ G_c $ a subset of $ G $ obtained by excluding all discrete and composite elements from the latter. For example, such procedure corresponds to excluding the inversion and SCT from the CG. Further, in order to avoid going into undesirable mathematical details, it will be \textit{required} that $ G_c $ is a group, that $ P_\mu \in G_c $, and that the algebra of $ G_c $, $ AG_c $, is homogeneously reductive with respect to the decomposition 
\begin{equation} \label{AlgDivis}
AG_c = P_\mu \oplus H_a \;,
\end{equation}
where $ H_a $ supplement $ P_\mu $ to the full set of generators of $ G_c $. As it will become clear shortly, such requirements are rather general. Then, the goal is to show that the CST is applicable for the construction of $ G $--invariant Lagrangians for fields $ \psi(x) $ living on $ A^d_{iso} $. 

To achieve the goal above, two technical statements need to be proved. The first one is that the action of $ H = \lbrace e^{iH_ab^a} \rbrace $ leaves not only $ \vec{0} $ invariant, but all $ z_m $ as well. Indeed, suppose otherwise. Then, as $ H $ is a continuous group, it is legitimate to consider infinitesimal transformations. If they act non--trivially on any of $ z_m $, then the latter is mapped to some ``ordinary'' point, which belongs to the orbit of $ \vec{0} $ under the action of $ e^{iP_\mu a^\mu} $ for some $ \vec{a} $. Therefore, $ z_m $ can be obtained by the action of $ e^{iP_\mu a^\mu} \in G_c $ on $ \vec{0} $. But this contradicts the condition that $ z_m $ is identified with $ T_m $ within isomorphism (\ref{IsomGen}), which finishes the proof. Similarly, it can be proved that $ T_m $ mixes $ \lbrace z_m \rbrace $ and $ \vec{0} $ only between each other. 

The second statement is that the group automorphisms 
\begin{equation} \label{AutoGroup}
W_m:~~ G \rightarrow G\,, ~~ \forall g \in G \rightarrow T_m\, g\, T_m^{-1}\;,
\end{equation}  
map $ H $ to itself. Indeed, for a given $ m $, automorphism (\ref{AutoGroup}) can be considered as the following isomorphism of $ A^d_{iso} $ to itself,
\begin{equation} \label{IsoHomSpace}
A^d_{iso} \rightarrow A^d_{iso}: ~~~ \forall\, \vec{v} \in  A^d_{iso} \rightarrow \hat{T}_m \vec{v},
\end{equation}
where $ \hat{T}_m $ is the representation of $ T_m $ acting on $ A^d_{iso} $. As it follows from the previous paragraph, (\ref{IsoHomSpace}) mixes $ \lbrace z_m \rbrace $ and $ \vec{0} $ only between each other. Then, since $ H $ is the stability group of  $ \lbrace z_m\,, 0 \rbrace  $, both of the isomorphisms above map $ H $ to itself, QED. 

From this result immediately follows that the full set of generators of $ G $ is
\begin{equation} \label{Algebra}
P_\mu \, , ~~~ K^{(m)}_\mu \equiv T_m \, P_\mu \, T_m^{-1} \,, ~~~ H_a\;.
\end{equation}
Moreover, as $ AG_c $ is homogeneously reductive with respect to decomposition (\ref{AlgDivis}), one has
\begin{equation} \label{ReductExtend}
[K^{(m)}_\mu , H] = T_m [P_\mu, H ] T^{-1}_m \subset K^{(m)}_\mu \;.
\end{equation}
The two relations above establish remarkable properties of the algebras of the groups with discrete elements, namely, the explicit form of a full set of generators and their commutation relations. In particular, (\ref{Algebra}) and (\ref{ReductExtend}) correctly reproduce the structure of the conformal algebra.

Now one can go back to the question of applying the CST to such groups. As it was explained in section \ref{sec:3}, the proper coset space endows $ A^d_{iso} $ with a natural atlas. Thus, in a general case one must use the extended, ``n--orbit'' coset space. As the stability group of all special points, up to their exchange, is $ H_n = H \times T_1 \times ... \times T_n $, the appropriate coset space is $ G/H_n $. In particular, from (\ref{ReductExtend}) follows that $ G/H_n $ is homogeneously reductive, and thus the CST is applicable. Before writing down the coset space, one needs to choose either $ P_\mu $ or any of $ K^{(m)}_\mu $ as the generators of ``translations''. Suppose that $ P_\mu $ were chosen for that role, then the proper coset space reads
\begin{equation} \label{GenCoset}
g_H = e^{iP_\mu x^\mu} e^{iK^{(1)}_{\mu_1} y_{(1)}^{\mu_1}} ... e^{iK^{(n)}_{\mu_n} y_{(n)}^{\mu_n}}\;.
\end{equation} 
Importantly, (\ref{ReductExtend}) guarantees that the MCF $ \omega_P^\mu $ and $ \omega_{K^{(m)}}^\mu $ do not mix with each other under the action of $ H $, as it was noticed previously on the example of the CG. Also, it implies that $ K_\mu^{(m)} $ cannot be included into the Cartan algebra of $ G $, and thus matter fields do not carry a separate index characterizing their charge under the action of $ K_\mu^{(m)} $. Consequently, the action of $ e^{K^{(m)}_\mu b^\mu} $ on fields is realized as the left action of $ G $ on $ G/H $, which invokes field transformation (\ref{IndActonMatter}) with parameters depending on $ x^\mu $ and $ b^\mu $, but not on $ y^\mu_{(m)} $, as guaranteed by (\ref{ReductExtend}).

Following the same logic as for the CG, the method of induced representations and the group automorphisms require $ d y^\mu_{(m)} $ to be the pullbacks of $ dx^\mu $ after the corresponding change of coordinates. This fixes $ y^\mu_{(m)} $ to obey the gluing maps,
\begin{equation} \label{GenGlue}
y^\mu_{(m)}(x) = \hat{T}_m x^\mu \;.
\end{equation}
In addition, since these relations are compatible with all symmetries, the same reasoning as for the CG applies for arguing that they are obtained by setting the homogeneously transforming forms $ \omega^\nu_{K^{(m)}} $ to zero. Consequently, (\ref{GenGlue}) will always be solutions of the EqM of $ y^\mu_{(m)} $ only. 

Thus, all machinery of the CST is applicable to such groups. That is, to obtain the ingredients for the construction of $ G $--invariant Lagrangians, one calculates the corresponding Maurer--Cartan 1--forms and, if needed, introduces matter fields with the covariant derivatives defined via (\ref{CovarDerMatter}). Then, invariant actions are obtained as $ H $--invariant combinations of $ \psi,~ D\psi, ~ \omega^\nu_{K^{(m)}} $, and $ \omega_P^\mu $. The requirement for (\ref{GenGlue}) to be a solution of the EqM of $ y^\mu_{(m)} $ ensures that the discrete symmetries are indeed symmetries of a theory. 

Finally, the same reasoning as for the NGF for SCT applies for arguing that the breakdown of $ K_\mu^{(m)} $ does not give rise to dynamical NGF. Indeed, as all $ T_m $ are symmetries of a theory with a spontaneously broken $ G $, $ y^\mu_{(m)} $ is forced to provide the gluing maps between coordinate charts on the manifold, thus yielding them non--dynamical. The argument based on the polar decomposition can also be generalized as follows. Above it was shown that fields in such theories do not carry an index characterizing their charge under the action of $ K_\mu^{(m)} $. Hence, the action of the latter on the fields is expressible in terms of the action of $ P_\mu $ and $ H_a $, likewise the action of SCT reduces to that of translations, dilation and Lorentz transformation. Then, the polar decomposition demonstrates that there is no need in introducing the NGF for $ K_\mu^{(m)} $.

\section{Conclusion}
\label{sec:6}

In the paper, the method of applying the CST to the construction of conformally--invariant Lagrangians was developed. A careful handling of discrete symmetries was found to be a key to obtaining the proper interpretation of the NGF for SCT in an unbroken phase of a theory. This allowed to question the inverse Higgs phenomenon for the CG undergoing SSB. In more details, the main results of the paper are the following.

The presented construction clarifies the special role of the NGF for SCT and other similar non--linear spacetime symmetries. As it was demonstrated, they play a two--folded role: they are fields, but a solution of their EqM is fixed by the symmetries to provide the gluing map between the coordinate charts on the homogeneous space of a symmetry group. This ensures self--consistency of the theory by guaranteeing that discrete symmetries are indeed symmetries of the theory. The defining property of the correct coset space is that it endows the manifold with a natural atlas. As a consequence, one must use the extended, ``n--orbit'' coset space $ G/H_n $, where $ H_n $ is a stability group (up to exchange) of the origin and all points of the homogeneous space of $ G $ identified with the discrete elements. Further, the requirement for the proper gluing maps to be solutions of the EqM of such NGF invokes additional constraints on a theory. The theories satisfying these constraints are those that are indeed invariant under an action of the discrete elements of the symmetry group. For example, for the CG it forces the virial current to vanish or to be a total derivative, thus selecting the conformally--invariant theories from a bigger class of scale--invariant ones. 

In section \ref{sec:5} it was shown that the developed technique is applicable to groups whose homogeneous space is homogeneously reductive after the exclusion of all non--linear discrete symmetries, including composite ones. From the mathematical point of view, it is the existence of involute group automorphisms generated by discrete elements that guarantee applicability of the CST. Indeed, not only they fix the dependence of the ``NGF'' on the coordinates, but the structure of the algebra of such groups as well. 

Another finding was a special class of Lagrangians containing the interaction terms between the NGF for non--linear spacetime symmetries and matter fields only via a full derivative.  Dropping the total derivative term yields a new type of the Wess--Zumino term, which can arise on the manifolds whose atlas must include more than one coordinate chart. The CFT described by such class of Lagrangians are those whose virial current is a total derivative.

This work also questions the usage and interpretation of IHC. Although the usage of IHC allowed to reproduce the effective Lagrangian in subsection \ref{subsec:4-1}, this method contradicts the symmetry requirements and does not have a clear physical interpretation. On the other hand, the developed technique allowed to obtain the effective Lagrangian as well. Moreover, the way it treats the NGF for SCT in a broken phase of a theory has a clear interpretation -- the effective Lagrangian must inherit general properties of the initial one, for which constraints (\ref{VarySCT}) must be fulfilled. The author believes that the provided evidence is strong enough for concluding that imposing IHC is not the correct way of obtaining effective Lagrangians of such theories. It should be mentioned that the paper questions the usage of IHC as a proper way of working only with the NGF corresponding to composite symmetries, like SCT in the CG. Due to that reason, the obtained results are not applicable in other contexts, considered, for example, in \cite{Nicolis:2013sga,Endlich:2013vfa}.

By using the polar decomposition it was shown that the breakdown of SCT and similar symmetries does not give rise to dynamical NGF. This happens because the action the corresponding generators on fields is expressible in terms of the action of the other, basic generators ($ H_a $ in section \ref{sec:5}). Consequently, such generators do not produce independent fluctuations of an order parameter, and thus there is no need in introducing the associated NGF. The same was also demonstrated by applying the Nambu--Goldstone theorem for SSB of the CG down to the Poincare subgroup. Namely, in general case, the number of NGF equals the number of independent actions of the broken generators on the vacuum. Since SCT do not have their own action on fields, they never invoke dynamical NGF.

The results obtained in the paper can be applied to the construction of conformally invariant higher--spins theories in the AdS space \cite{Vasiliev:2003ev}. Also, the clarified role of the NGF for SCT may shed new light the AdS/CFT duality. Considering the plans of a future work, the author is going to elaborate on the usage and interpretation of IHC for SSB of spacetime symmetries in other contexts in the proceeding paper. As another important perspective, the results obtained in the paper should be explained from the perspectives of quantum field theory.

\vspace{0.2cm}
\textbf{Acknowledgements.} The author is thankful to E. Ivanov, S. Sibiryakov and A. Shkerin for useful discussions and comments on the draft of the paper. The work was supported by the Grant 14-22-00161 of the Russian Science Foundation.

\subsection*{Appendix}

Here it will be shown that Lagrangian (\ref{OnlyViaDer}) contains the interaction terms between $ \psi $ and $ y^\nu $ only via full derivative. Without loss of generality, $ \psi $ can be taken to be a vector field, since higher spin fields are formed as a tensor product of spin--1 representations\footnote{There is no need in considering spin--$\frac{1}{2}$ fields because the corresponding virial current, provided that the kinetic term is canonical, vanishes. Spin--$\frac{3}{2}$ fields would not be considered in the paper.}. For typographical convenience, the Latin letters will be used to denote the corresponding vector index of $ \psi_a $ when possible.  Then, assuming that the kinetic term is quadratic in $ \psi_a $, in general case it can be written as
\begin{equation} \label{Kinetic}
\begin{split}
&\frac{1}{2}C^{\mu a \nu b}(D_\mu \psi_a)(D_\nu \psi_b)  = \\ 
= \frac{1}{2} C^{\mu a \nu b} \big( \partial_\mu \psi_a \partial_\nu \psi_b + &2 \partial_\mu\psi_a y^\sigma (\hat{N}_{\sigma\nu}\psi_b) + y^\sigma (\hat{N}_{\sigma\nu} \psi_b) y^\rho (\hat{N}_{\rho\mu}\psi_a) \big) \;,
\end{split}
\end{equation}
where $ C^{\mu a \nu b} $ is some constant tensor, symmetric in $ (\mu a) \leftrightarrow (\nu b) $. If the virial current is a total derivative, from (\ref{Kinetic}) it follows that
\begin{equation} \label{KinSecond}
C^{\mu a \nu b}\partial_\mu\psi_a \hat{N}_{\rho\nu}\psi_b = \partial_\mu L^\mu_\rho ~~\Rightarrow ~~ \frac{\delta L^\mu_\rho}{\delta\psi_a} = C^{\mu a \nu b} \hat{N}_{\rho\nu}\psi_b \;.
\end{equation}
To proceed further, an explicit form of $ L^\mu_\sigma $ should be used. For a spin--1 field, its most general form is given by
\begin{equation} \label{DerForm}
L^\mu_\sigma = \alpha \psi^2 \delta^\mu_\sigma + \beta \psi^\mu\psi_\sigma\;,
\end{equation} 
where $ \alpha $ and $ \beta $ are some constants. In particular, note that the second term in Lagrangian (\ref{OnlyViaDer}) is allowed in the coset space framework, since it is a $ (SO(1,d)\times D) $--invariant combination of $ \psi_a,~ \omega_K^\nu $, and $ \omega_P^\mu $. Substituting (\ref{DerForm}) into (\ref{Kinetic}) allows to rewrite the third term therein in the form
\begin{equation} \label{KinThird}
\frac{1}{2} C^{\mu a \nu b} y^\rho y^\sigma (\hat{N}_{\rho\mu}\psi_a) (\hat{N}_{\sigma\nu}\psi_b) = \Delta\alpha y^2\psi^2 + \frac{\beta}{2} \left( y^2\psi^2 + (2\Delta - d)(y_\mu \psi^\mu)^2 \right)\;.
\end{equation}  
Finally, substituting (\ref{DerForm}) into the last term in (\ref{OnlyViaDer}) gives
\begin{equation} \label{AddSCT}
\epsilon_{\mu_0 ... \mu_d} L^{\mu_0}_\nu \omega_K^\nu \wedge \omega_P^{\mu_1} \wedge ... \wedge \omega_P^{\mu_d} = dy^\rho \wedge \tilde{L}_\rho + (2L^\mu_\rho y^\rho y_\mu - y^2 L^\mu_\mu)d^dx\;,
\end{equation}
where $ \tilde{L}_\rho $ is a differential form such that $ \partial_\mu L^\mu_\rho = d \tilde{L}_\rho $. Full Lagrangian (\ref{OnlyViaDer}) is a sum of (\ref{Kinetic}) and (\ref{AddSCT}), which, as it follows from (\ref{KinSecond}) and (\ref{KinThird}), contains $ y^\nu(x) $ only via full derivative, $ d(y^\rho \tilde{L}_\rho) $. 

\bibliography{cftcoset}

\end{document}